RESEARCH ARTICLE

# Rate Control Management of Atrial Fibrillation: May a Mathematical Model Suggest an Ideal Heart Rate?

Matteo Anselmino[1], Stefania Scarsoglio[2], Carlo Camporeale[3], Andrea Saglietto[1], Fiorenzo Gaita[1], Luca Ridolfi[3]*

1 Division of Cardiology, Department of Medical Sciences, "Città della Salute e della Scienza" Hospital, University of Turin, Turin, Italy, 2 DIMEAS -Department of Mechanical and Aerospace Engineering-, Politecnico di Torino, Turin, Italy, 3 DIATI -Department of Environmental, Land and Infrastructure Engineering-, Politecnico di Torino, Turin, Italy

* luca.ridolfi@polito.it

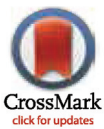

## Abstract

### Background
Despite the routine prescription of rate control therapy for atrial fibrillation (AF), clinical evidence demonstrating a heart rate target is lacking. Aim of the present study was to run a mathematical model simulating AF episodes with a different heart rate (HR) to predict hemodynamic parameters for each situation.

### Methods
The lumped model, representing the pumping heart together with systemic and pulmonary circuits, was run to simulate AF with HR of 50, 70, 90, 110 and 130 bpm, respectively.

### Results
Left ventricular pressure increased by 57%, from 33.92±37.56 mmHg to 53.15±47.56 mmHg, and mean systemic arterial pressure increased by 27%, from 82.66±14.04 mmHg to 105.3±7.6 mmHg, at the 50 and 130 bpm simulations, respectively. Stroke volume (from 77.45±8.50 to 39.09±8.08 mL), ejection fraction (from 61.10±4.40 to 39.32±5.42%) and stroke work (SW, from 0.88±0.04 to 0.58±0.09 J) decreased by 50, 36 and 34%, at the 50 and 130 bpm simulations, respectively. In addition, oxygen consumption indexes (rate pressure product – RPP, tension time index per minute – TTI/min, and pressure volume area per minute – PVA/min) increased from the 50 to the 130 bpm simulation, respectively, by 186% (from 5598±1939 to 15995±3219 mmHg/min), 56% (from 2094±265 to 3257±301 mmHg s/min) and 102% (from 57.99±17.90 to 117.4±26.0 J/min). In fact, left ventricular efficiency (SW/PVA) decreased from 80.91±2.91% at 50 bpm to 66.43±3.72% at the 130 bpm HR simulation.

## Conclusion

Awaiting compulsory direct clinical evidences, the present mathematical model suggests that lower HRs during permanent AF relates to improved hemodynamic parameters, cardiac efficiency, and lower oxygen consumption.

## Introduction

Atrial fibrillation (AF), the most common sustained tachyarrhythmia, affects 1% to 2% of the general population [1]. In case the arrhythmia progresses to permanent [2], on top of oral anticoagulants, rate control is recommended to reduce symptoms and improve quality of life [3]. Despite the routine prescription of AF rate control therapy [4], clinical evidence demonstrating a heart rate target is lacking. Only the RACE II clinical trial [5, 6] suggested that lenient and strict rate control strategies may not differ in terms of mid-term cardiovascular outcomes.

Awaiting further clinical evidences, the use of a novel mathematical model [7], validated through systematic comparison with directly measured parameters and able to simulate the response of the cardiovascular system in sinus rhythm and AF, holds the potential to theoretically suggest an optimal heart rate to target.

Aim of the present study was to run the mathematical model during simulated AF episodes with heart rate ranging from 50 to 130 bpm and predict hemodynamic parameters for each situation.

## Materials and Methods

### Mathematical model

The present lumped model, consisting of a network of compliances, resistances, and inductances, simulates the pumping heart together with the systemic and pulmonary circuits [7]. The resultant differential system is composed, for each cardiac chamber or vascular section, by an equation for the mass conservation, an equation of motion, and a linear state equation, and is expressed in terms of pressure, P, volume, V, and flow rate, Q. The differential equations are numerically solved through a multistep adaptative solver for stiff problems, based on the ode15s Matlab function [8]. All hemodynamic parameters and plots are computed and implemented in Matlab, as well. The model has been compared with more than thirty different clinical state of the art studies, providing an overall good agreement in predicting the impact of AF [7].

Both atria are maintained passive to mimic the loss of atrial kick, while RR values are extracted from an exponentially modified Gaussian distribution [7], which is the most common RR distribution recorded during AF [9]. By varying the HR from 50 to 130 bpm, each distribution is built keeping the coefficient of variation, $c_v = \sigma/\mu$ (03C3: standard deviation, $\mu$: mean of the RR distribution), equal to 0.24, which is the typical value observed during AF beating [10]. The resulting probability distribution functions for 50, 70, 90, 110, and 130 bpm, are shown in Fig. 1.

For every HR, 5000 cardiac cycles are computed, which allows the statistical stationarity of the modeling results. Therefore, all the variables of the present work are intended as averaged over 5000 periods.

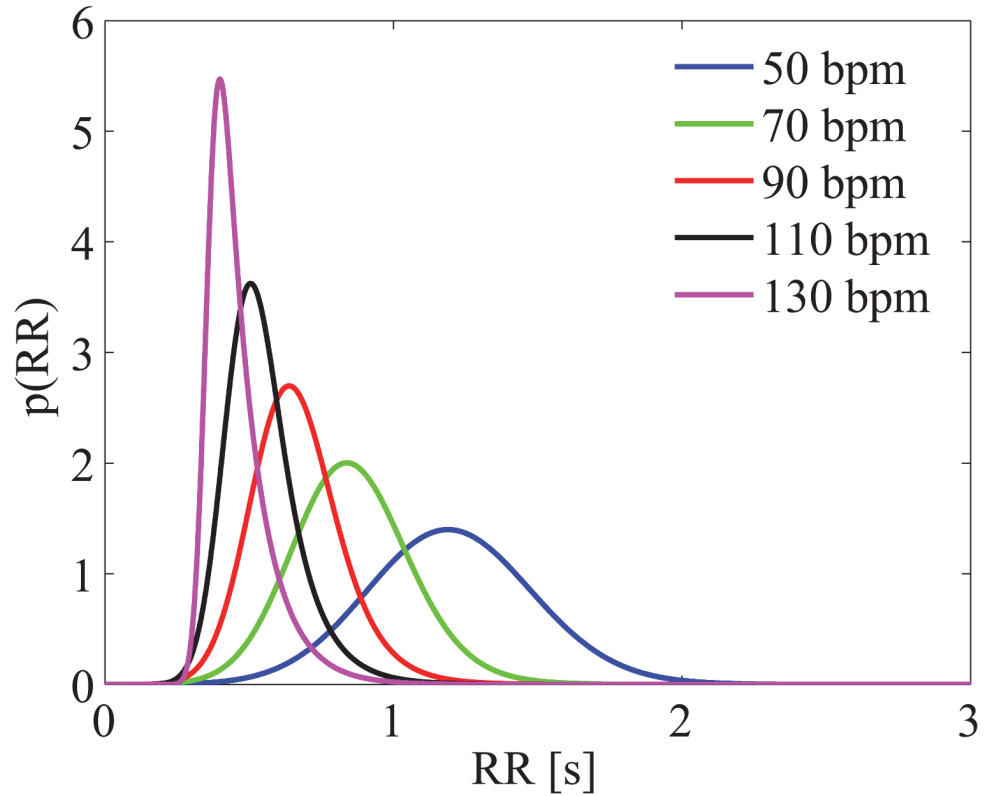

**Fig 1. RR distributions.** Probability distribution functions of RR interval for the different simulations are reported.

doi:10.1371/journal.pone.0119868.g001

### Definitions of variables

In terms of pressure and volume, by evaluating also end-systolic (es) and end-diastolic (ed) values, as well as left ventricular pressure peak values (maximum and minimum), the following parameters are computed: left atrial pressure ($P_{la}$, $P_{laed}$, $P_{laes}$), left atrial volume ($V_{la}$, $V_{laed}$, $V_{laes}$), left ventricular pressure ($P_{lv}$, $P_{lved}$, $P_{lves}$, $P_{lv,max}$, $P_{lv,min}$), left ventricular volume ($V_{lv}$, $V_{lved}$, $V_{lves}$), systemic arterial pressure ($P_{sas}$, $P_{sas,syst}$, $P_{sas,dias}$), pulmonary arterial ($P_{pas}$, $P_{pas,syst}$, $P_{pas,dias}$) and venous ($P_{pvn}$) pressures. End-systolic values refer to the instant defined by the closure of the aortic valve, while end-diastolic values correspond to the closure of the mitral valve.

Concerning left ventricle performance the following parameters are also computed: stroke volume, $SV = V_{lved} - V_{lves}$, ejection fraction, $EF = SV/V_{lved} \times 100$, stroke work, SW, evaluated as the area within the left ventricle pressure-volume loop, and cardiac output, $CO = SV \times HR$. To estimate the oxygen consumption, the following indirect measurements were computed [11]: rate pressure product, $RPP = P_{sas,syst} \times HR$, tension time index per minute [12], $TTI/min = P_{lv,mean} \times RR \times HR$, and pressure volume area per minute [13], $PVA/min = (PE + SW) \times HR$, where $PE = P_{lves} \times (V_{lves} - V_{lv,un})/2 - P_{lved} \times (V_{lved} - V_{lv,un})/4$ is the elastic potential energy ($V_{lv,un} = 5$ ml is the unstressed left ventricle volume), while SW is the stroke work. The left ventricular efficiency is defined by the ratio SW/PVA.

### Results

The mathematical model was run to simulate AF with heart rate (HR) of 50, 70, 90, 110 and 130 bpm, respectively. All computed parameters, stratified by HR, are listed in Table 1. Left

Table 1. Mean and standard deviation of computed parameters stratified by each simulation.

| Parameters | Results of simulations | | | | | Maximum % variation[a] |
|---|---|---|---|---|---|---|
| | 50 bpm | 70 bpm | 90 bpm | 110 bpm | 130 bpm | |
| $P_{la}$ [mmHg] | 9.81 ± 0.81 | 9.39 ± 0.77 | 9.16 ± 0.76 | 9.07 ± 0.76 | 9.08 ± 0.77 | -8 |
| $P_{laes}$ [mmHg] | 11.00 ± 0.35 | 10.41 ± 0.25 | 10.08 ± 0.17 | 9.91 ± 0.13 | 9.85 ± 0.13 | -10 |
| $P_{laed}$ [mmHg] | 10.10 ± 0.21 | 9.75 ± 0.14 | 9.59 ± 0.09 | 9.53 ± 0.10 | 9.53 ± 0.09 | -6 |
| $V_{la}$ [ml] | 62.76 ± 5.41 | 59.91 ± 5.16 | 58.42 ± 5.04 | 57.83 ± 5.07 | 57.88 ± 5.17 | -8 |
| $V_{laes}$ [ml] | 70.68 ± 2.31 | 66.72 ± 1.68 | 64.52 ± 1.11 | 63.40 ± 0.88 | 63.00 ± 0.84 | -11 |
| $V_{laed}$ [ml] | 64.68 ± 1.41 | 62.35 ± 0.95 | 61.26 ± 0.62 | 60.89 ± 0.65 | 60.89 ± 0.63 | -6 |
| **$P_{lv}$ [mmHg]** | 33.92 ± 37.56 | 40.16 ± 42.46 | 45.38 ± 45.42 | 49.69 ± 46.98 | 53.15 ± 47.56 | **+57** |
| $P_{lves}$ [mmHg] | 91.75 ± 3.92 | 95.74 ± 2.78 | 96.86 ± 2.25 | 96.51 ± 2.09 | 95.46 ± 1.84 | +6 |
| **$P_{lved}$ [mmHg]** | 15.46 ± 0.75 | 16.32 ± 0.97 | 17.24 ± 1.08 | 18.21 ± 1.14 | 19.08 ± 1.02 | **+23** |
| $P_{lv,max}$ [mmHg] | 103.9 ± 5.1 | 111.1 ± 3.9 | 115.1 ± 2.6 | 117.0 ± 1.8 | 117.5 ± 1.5 | +13 |
| $P_{lv,min}$ [mmHg] | 4.85 ± 0.11 | 4.77 ± 0.06 | 4.79 ± 0.05 | 4.82 ± 0.05 | 4.84 ± 0.04 | 0 |
| **$V_{lv}$ [ml]** | 101.4 ± 33.1 | 93.72 ± 28.97 | 88.50 ± 25.79 | 84.24 ± 23.27 | 80.53 ± 21.15 | **-21** |
| **$V_{lves}$ [ml]** | 48.94 ± 2.88 | 53.53 ± 2.76 | 56.70 ± 2.15 | 58.54 ± 1.54 | 59.35 ± 1.16 | **+21** |
| **$V_{lved}$ [ml]** | 126.4 ± 5.7 | 117.2 ± 6.1 | 110.1 ± 6.5 | 103.9 ± 7.2 | 98.45 ± 7.11 | **-22** |
| **$P_{sas}$ [mmHg]** | 82.66 ± 14.04 | 93.35 ± 11.65 | 99.72 ± 9.83 | 103.4 ± 8.5 | 105.3 ± 7.6 | **+27** |
| **$P_{sas,dias}$ [mmHg]** | 64.99 ± 8.90 | 78.20 ± 7.47 | 86.46 ± 5.88 | 91.73 ± 4.74 | 94.92 ± 3.93 | **+46** |
| $P_{sas,syst}$ [mmHg] | 103.8 ± 5.1 | 111.0 ± 3.9 | 115.0 ± 2.6 | 116.9 ± 1.8 | 117.4 ± 1.5 | +13 |
| $P_{pas}$ [mmHg] | 18.72 ± 5.03 | 19.62 ± 4.25 | 20.18 ± 3.68 | 20.54 ± 3.25 | 20.78 ± 2.96 | +11 |
| **$P_{pas,dias}$ [mmHg]** | 12.67 ± 1.26 | 14.05 ± 1.38 | 15.18 ± 1.32 | 16.06 ± 1.19 | 16.73 ± 1.08 | **+32** |
| $P_{pas,syst}$ [mmHg] | 27.43 ± 0.95 | 26.51 ± 0.77 | 25.96 ± 0.68 | 25.58 ± 0.63 | 25.31 ± 0.60 | -8 |
| $P_{pvn}$ [mmHg] | 10.20 ± 0.65 | 9.83 ± 0.56 | 9.64 ± 0.49 | 9.57 ± 0.44 | 9.59 ± 0.41 | -6 |
| **SV [ml]** | 77.45 ± 8.50 | 63.61 ± 8.84 | 53.39 ± 8.43 | 45.32 ± 8.54 | 39.09 ± 8.08 | **-50** |
| **EF [%]** | 61.08 ± 4.40 | 54.03 ± 5.22 | 48.19 ± 5.27 | 43.24 ± 5.70 | 39.32 ± 5.42 | **-36** |
| **SW [J]** | 0.88 ± 0.04 | 0.81 ± 0.07 | 0.73 ± 0.08 | 0.65 ± 0.10 | 0.58 ± 0.09 | **-34** |
| **CO [l/min]** | 4.02 ± 0.66 | 4.57 ± 0.58 | 4.88 ± 0.47 | 5.03 ± 0.43 | 5.11 ± 0.35 | **+27** |
| **RPP [mmHg/min]** | 5598 ± 1939 | 8330 ± 2612 | 10977 ± 2953 | 13601 ± 3292 | 15995 ± 3219 | **+186** |
| **TTI/min [mmHg s/min]** | 2094 ± 265 | 2479 ± 304 | 2793 ± 310 | 3053 ± 314 | 3257 ± 301 | **+56** |
| **PVA/min [J/min]** | 57.99 ± 17.90 | 78.82 ± 20.66 | 95.17 ± 22.41 | 107.9 ± 24.7 | 117.4 ± 26.0 | **+102** |
| **SW/PVA [%]** | 80.91 ± 2.91 | 76.42 ± 3.43 | 72.61 ± 3.50 | 69.20 ± 4.05 | 66.43 ± 3.72 | **-18** |

$P_{la}$, left atrium pressure; $P_{laes}$, left atrium end-systolic pressure; $P_{laed}$, left atrium end-diastolic pressure; $V_{la}$, left atrium volume; $V_{laes}$, left atrium end-systolic volume; $V_{laed}$, left atrium end-diastolic volume; $P_{lv}$, left ventricular pressure; $P_{lves}$, left ventricular end-systolic pressure; $P_{lved}$, left ventricular end-diastolic pressure; $P_{lv,max}$, left ventricular maximum pressure; $P_{lv,min}$, left ventricular minimum pressure; $V_{lv}$, left ventricular volume; $V_{lves}$, left ventricular end-systolic volume; $V_{lved}$, left ventricular end-diastolic volume; $P_{sas}$, mean systemic arterial pressure; $P_{sas,dias}$, diastolic systemic arterial pressure; $P_{sas,syst}$, systolic systemic arterial pressure; $P_{pas}$, mean pulmonary arterial pressure; $P_{pas,dias}$, diastolic pulmonary arterial pressure; $P_{pas,syst}$, systolic pulmonary arterial pressure; $P_{pvn}$, pulmonary vein pressure; SV, stroke volume; EF, ejection fraction; SW, stroke work; CO, cardiac output; RPP, rate pressure product; TTI/min, tension time index per minute; PVA/min, pressure volume area per minute; SW/PVA, left ventricular efficiency.

[a] parameters showing a maximum % variation above ±15% with respect to reference (50 bpm) are reported in bold.

doi:10.1371/journal.pone.0119868.t001

ventricular pressure ($P_{lv}$), systemic pressure ($P_{sas}$) and cardiac mechano-energetic indexes (e.g. SV, EF, SW and ventricular efficiency) varied more than 15% within the different HR simulations.

Mean value and representative examples of temporal series of left ventricular pressure are shown in Fig. 2: $P_{lv}$ increased by 57%, from 33.92±37.56 mmHg to 53.15±47.56 mmHg at the 50 and 130 bpm simulations, respectively.

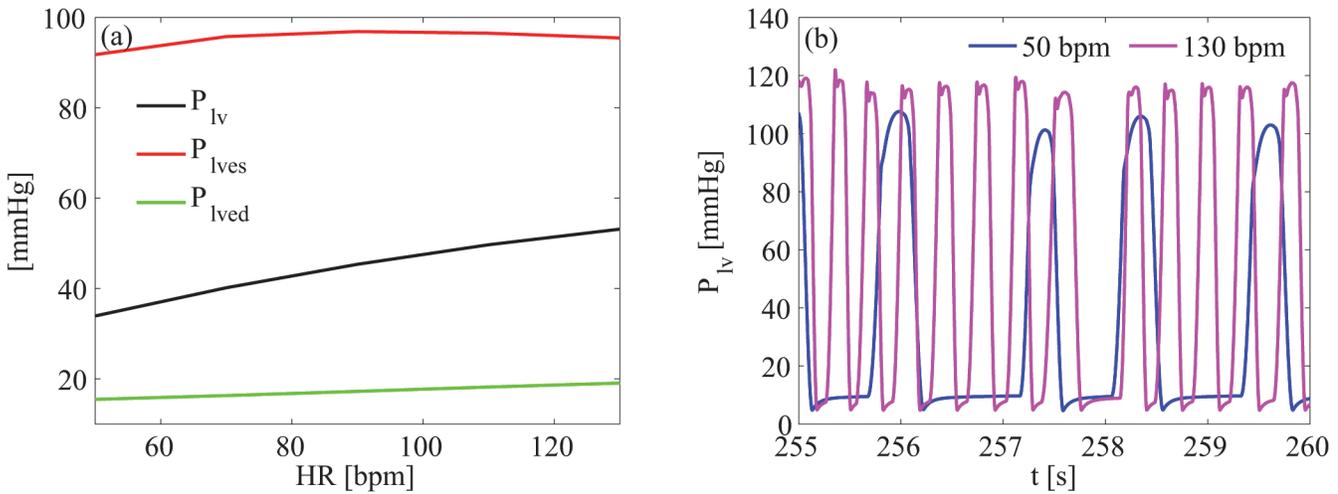

**Fig 2. Left ventricular pressure.** (a) mean left ventricular pressure as function of heart rate; (b) representative left ventricular pressure time series of 50 and 130 bpm simulations. $P_{lv}$, left ventricular pressure; $P_{lves}$, left ventricular end-systolic pressure; $P_{lved}$, left ventricular end-diastolic pressure.

doi:10.1371/journal.pone.0119868.g002

Systemic pressure variations are illustrated in Fig. 3. Mean $P_{sas}$ increased by 27%, from 82.66±14.04 mmHg to 105.3±7.6 mmHg at the 50 and 130 bpm simulations, respectively. In details, systolic pressure ($P_{sas,syst}$) shifted from 103.8±5.1 mmHg (50 bpm) to 117.4±1.5 mmHg (130 bpm), and diastolic pressure ($P_{sas,dias}$) from 64.99±8.90 mmHg (50 bpm) to 94.92±3.93 mmHg (130 bpm).

Eventually, heart performance was assessed by several mechanic and energetic parameters (Fig. 4). Concerning left ventricle mechanics, SV (from 77.45±8.50 to 39.09±8.08 mL), EF (from 61.10±4.40 to 39.32±5.42%) and SW (from 0.88±0.04 to 0.58±0.09 J) decreased by 50, 36 and 34%, at the 50 and 130 bpm simulations, respectively. In addition, oxygen consumption indexes (RPP, TTI/min and PVA/min) increased from the 50 to the 130 bpm simulation, respectively, by 186% (from 5598±1939 to 15995±3219 mmHg/min), 56% (from 2094±265 to 3257±301 mmHg s/min) and 102% (from 57.99±17.90 to 117.4±26.0 J/min). In fact, left

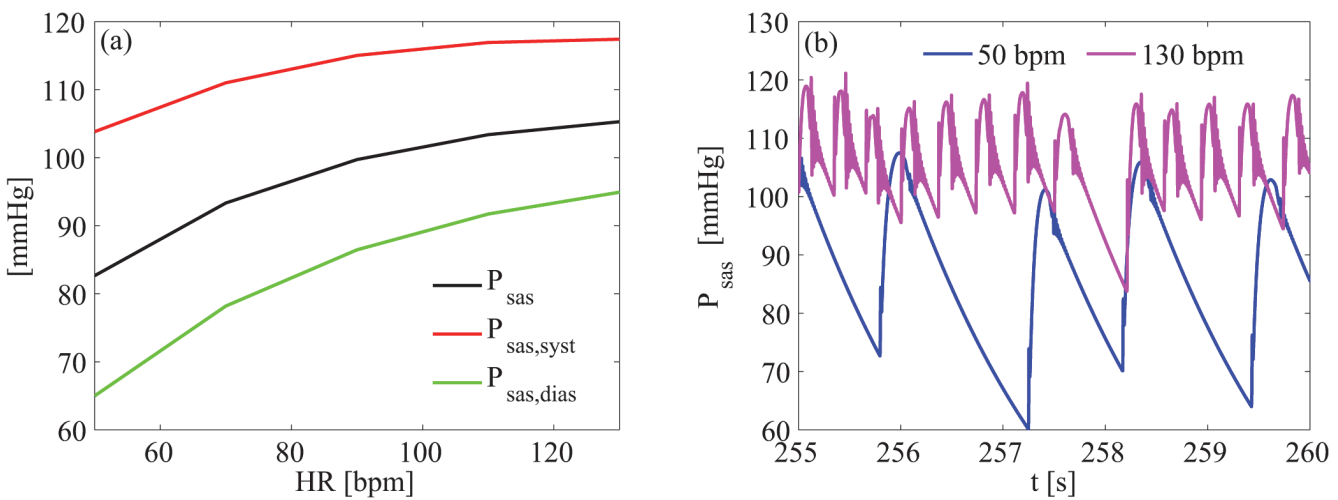

**Fig 3. Systemic arterial pressure.** (a) mean values of systemic arterial pressure as function of heart rate; (b) representative systemic arterial pressure time series of 50 and 130 bpm simulations. $P_{sas}$, mean systemic arterial pressure; $P_{sas,dias}$, diastolic systemic arterial pressure; $P_{sas,syst}$, systolic systemic arterial pressure.

doi:10.1371/journal.pone.0119868.g003

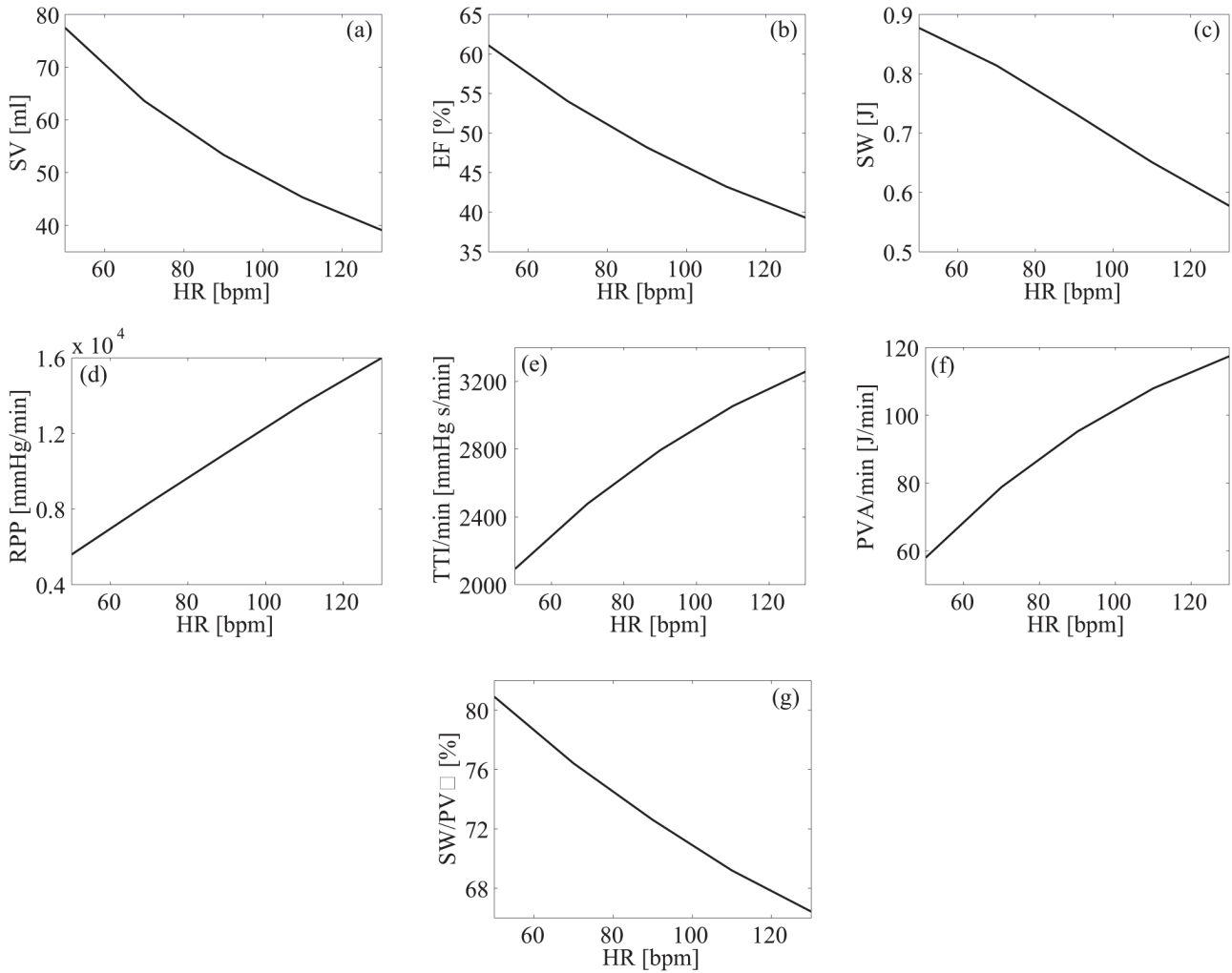

**Fig 4. Mechanic and energetic indexes of left heart.** Mean values of mechanic and energetic indexes are plotted as function of heart rate. (a) stroke volume, SV; (b) ejection fraction, EF; (c) stroke work, SW; (d) rate pressure product, RPP; (e) tension time index per minute, TTI/min; (f) pressure volume area per minute, PVA/min; (g) left ventricular efficiency, SW/PVA.

doi:10.1371/journal.pone.0119868.g004

ventricular efficiency (SW/PVA) decreased from 80.91±2.91% at 50 bpm to 66.43±3.72% at the 130 bpm HR simulation.

## Discussion

Based on the presented mathematical simulations a slower (50 bpm), compared to higher (130 bpm), HR during AF relates to improved ventricular pressure, systemic pressure and left ventricular efficiency (e.g. SW/PVA).

Given several clinical trial (AFFIRM [14, 15], RACE [16]) proving, within patients with persistent AF, that rate is not inferior to rhythm control in terms of mortality from cardiovascular causes, AF rate control therapy is widely prescribed. Despite this, clinical evidence demonstrating a clear heart rate target is lacking. A randomized multicenter non-inferiority clinical trial, the RACE II [5, 6], suggested that, in patients with permanent AF, lenient rate control (target resting HR below 110 bpm) is not inferior to strict (target resting HR below 80 and during moderate exercise below 110 bpm) in terms of cardiovascular outcomes. Moreover, substudies of this trial also underlined that stringency of rate control does not influence neither cardiac remodeling [17] nor quality of life [18].

However, as previously highlighted [19], the RACE II study presented several limitations. First, the modest difference in average heart rates achieved in the lenient and strict control groups (85 and 75 bpm, respectively) and the limited sample size of the study (about 300 patients per group). In addition, the number of patients who met the primary outcome of the study (a composite of death from cardiovascular causes, hospitalization for heart failure, stroke, systemic embolism, major bleeding, arrhythmic events, cardiac arrest, life-threatening adverse effects of rate-control drugs and implantation of a pacemaker or cardioverter–defibrillator) was small, most probably due to the low-risk population enrolled (e.g. patients with previous stroke were excluded).

Awaiting mandatory further clinical evidences, the present study aims at investigating, with a model-based approach, the global response of the cardiovascular system during episodes of AF at different ventricular rates. The present mathematical model has previously been thoroughly validated and showed strong concordance of the computed parameters with several data directly measured in vivo [7]. However, two considerations should be kept in mind: first, the model predicts hemodynamic effects of AF in absence of other associated conditions or pathologies, e.g. hypertension [20], that could themselves affect cardiovascular parameters; second, the model does not consider the impact that rate control drugs (e.g. digoxin, beta blockers, non-dihydropiridine calcium channel blockers) could exert on the cardiovascular system. As a consequence, the reported parameters differ among each simulation strictly depending on ventricular rate.

Left ventricular pressure greatly increased at faster ventricular rates (as shown in Fig. 2). The relative shortening of diastolic time at the fastest HR simulations, in our opinion, mainly drives this finding. In fact, the increase in mean $P_{lved}$ values (below 4 mmHg) does not balance the dramatic absolute and relative increase of the mean $P_{lv}$ (nearly 20 mmHg). Moreover, $P_{lves}$ and both pressure peak values do not significantly vary (minimum pressure peaks remain even constant as HR increases).

Contextually, it is important to underline that the reduction in $V_{lv}$ is mainly founded on the decrease of $V_{lved}$. In general, ventricular filling is known to be reduced during AF, due to loss of the atrial kick, however, the contribution of this cardiac phase is highly dependent on heart rate, becoming fundamental in case it is increased (e.g. physical exercise [21]), due to the fact that passive ventricular filling in early diastole is reduced concomitantly to a shortened diastolic time. The same mechanisms may, in fact, explain the greater $V_{lved}$ reduction in AF simulations with a faster ventricular rate. In addition higher HRs lead also to a quite strong rise of $P_{sas}$, most probably related to an increase in diastolic pressure accounting for a greater afterload for the left heart to overcome.

Based on the model's predictions, lower HRs during AF relate to improved mechanic and energetic indexes. First, the reduction of SV and SW seen with the progressively higher HRs translates into a left ventricle's EF decrease by far greater than the physiological reduction expected [22]. Furthermore, the increase in oxygen consumption indexes is relevant with the progressively higher HRs simulations. In fact, RPP triplicates and PVA/min doubles their mean values from the lowest (50 bpm) to the highest (130 bpm) HR simulation, extensively supporting that faster ventricular rates relate to a rise in cardiac energetic expense not associated with a concomitant increase in mechanical performance (e.g. CO). Finally, left ventricle efficiency (SW/PVA) proved to significantly decrease at progressively higher HRs simulated during the AF episodes, well synthesizing how mechanic and energetic indexes of the heart present an improved profile at slower HRs that allow the fibrillating heart to better convert energy into external work.

Eventually, although the present model currently does not present the ability to predict such setting, during exertion (a situation in which inotropism and chronotropism increase due to sympathetic stimulation) the reduction in left ventricle efficiency caused by faster HR may become even more limiting, perhaps, at least partially, accounting for the lower resistance to effort common in subjects with AF.

## Limitations

In addition to what previously discussed, the following limitations must be taken in account. First, the present mathematical model simulates a denervated heart; the effects of the autonomic nervous system on cardiac performance are therefore not included. Second, due to difficulties in the mathematical modelling of the coronary arteries dynamics, in the present model the coronary circle is not taken in account. Third, the present model-based approach predicts the global response of the cardiovascular system during episodes of AF at different HRs. For the purpose of the present study we considered "relevant" variations (maximum % variation), within the different HR simulations, above 15%. Concerning parameters not reaching this limit, such as left atrial parameters ($P_{la}$ and $V_{la}$) or pulmonary pressures ($P_{pas}$ and $P_{pvn}$) it cannot be concluded if either different HRs do not significantly affect their values, or the present model may not be able to detect variations.

## Conclusion

In conclusion, awaiting compulsory direct clinical evidences, the present mathematical model suggests that lower HRs during permanent AF relates to improved hemodynamic parameters and cardiac efficiency, resulting in lower oxygen consumption for a given cardiac work.

## Author Contributions

Conceived and designed the experiments: MA AS FG LR. Performed the experiments: SS CC LR. Analyzed the data: AS MA. Contributed reagents/materials/analysis tools: SS CC LR. Wrote the paper: MA AS SS CC FG LR.